\documentstyle[twocolumn,psfig,rotating]{mn}

\newif\ifAMStwofonts

\def\lsim{\lower.5ex\hbox{$\; \buildrel < \over \sim \;$}}
\def\gsim{\lower.5ex\hbox{$\; \buildrel > \over \sim \;$}}

\title{Can ambipolar diffusion solve the cosmic lithium problem?}
\author[L. Chuzhoy]
       {Leonid Chuzhoy \\
        McDonald Observatory and Department of Astronomy, The University of Texas at Austin, RLM 16.206, Austin, TX 78712, USA\\
	email: chuzhoy@astro.as.utexas.edu}

\pagerange{\pageref{firstpage}--\pageref{lastpage}}
\begin{document}
\maketitle

\begin{abstract}
Recent WMAP observations indicate that the primordial abundance of lithium should have been 2-4 times higher than the value measured in low-metallicity stars. We show that this discrepancy can be explained by the process of ambipolar diffusion in protostellar clouds, which leads to depletion of elements with low ionization potential, such as lithium and sodium. In high-metallicity stars, [Fe/H]$\gsim -1$, the depletion of lithium due to this process is found to be negligible, but for  [Fe/H]$\lsim -1.5$ the lithium abundance falls by a factor of $\sim 3$, consistent with WMAP observations.
\end{abstract}


\section{\label{Int}Introduction}

Since the discovery of a nearly constant lithium abundance \cite{Sp} in the metal poor stars, there has been a tendency to assume that the measured value reflects the primordial abundance. However, in recent years this interpretation came into conflict with the high value of $\Omega_{\rm b}$ obtained by the WMAP satellite \cite{Sp3,Sp6}, which in the standard big bang nucleosynthesis model corresponds to a factor of 2-4 larger lithium abundance.
In this paper we show that the process of ambipolar diffusion in protostellar clouds can provide a natural explanation to the missing lithium problem. 

The process of ambipolar diffusion, i.e., an outflow of charged particles from neutral gas caused by magnetic force, is primarily thought of as a way to reduce the strength of magnetic field. It has been long recognized that in addition to efficient cooling, which removes the thermal pressure support, the collapse of the diffuse gas into a star requires some mechanism to reduce the magnetic pressure, for which purpose ambipolar diffusion seems to be the most promising candidate \cite{MS,Na,MP,MTK}. 
As we show in this paper, besides reducing the magnetic flux, ambipolar diffusion can also affect the chemical composition of the gas by preferentially expelling elements with low ionization potential. In particular, we find that in the metal-poor stars ([Fe/H]$\lsim -1.5$) \footnote{[X/H]$=Log_{10}(n(X)/n(H))- Log_{10}(n(X)/n(H))_{\odot}$} the abundance of lithium falls by a factor of $\sim 3$, which allows to explain the discrepancy with the WMAP results.

The paper is organized as follows. In \S 2, we discuss how ionization balance evolves in a collapsing protostellar cloud. In \S 3, we estimate how the chemical composition of the cloud changes during the collapse. In \S 4, we summarize our results.

\section{Ionization balance}
Partial ionization of protostellar clouds can be sustained by several different processes. Ionizations by cosmic rays, which are believed to be dominant at the present epoch, initially produce mostly hydrogen and helium ions. However, since the charge-transfer reactions are typically much faster than radiative recombinations, the light elements pass the charge to metals with lower ionization potential, which therefore make up most of the ions \cite{OD}. In equilibrium the ionizations by cosmic-rays are balanced by recombinations
\begin{eqnarray}
\label{cosm}
\zeta n=\alpha x^2 n^2,
\end{eqnarray}
where $\zeta$ is the cosmic rays intensity, $\alpha$ is the recombination coefficient, $n$ is the gas particles number density and $x$ is the fractional ionization. Neglecting the possible change in $\alpha$, which may result from change in temperature and chemical composition of the gas, we find from eq. (\ref{cosm}) that the fractional ionization scales with density as $x\propto n^{-1/2}$. 

At high redshifts, when the UV flux was significantly higher and the cosmic-ray flux may have been lower, ionizations by UV photon with energies below 13.6 eV (hydrogen ionization potential) to which the neutral hydrogen is transparent, may have been the dominant source of ions in protostellar clouds. In the optically thin limit $x$ is again proportional to $n^{-1/2}$. At high densities the dependence of $x$ on $n$ becomes non-trivial, as metals with high ionization potential become self-shielded, while those with low ionization potential (which can be ionized by wider spectral range) are still optically thin. Nevertheless, up to the point when all abundant metals become self-shielded, the dependence of $x$ on $n$ should be only slightly steeper than in the optically thin limit.

\section{Protostellar cloud collapse}
We consider the collapse of a protostellar cloud under its own gravity, starting from a point of magnetostatic equilibrium. Following previous authors \cite{CF}, we ignore the detailed structure of a cloud, assuming instead that it can be described as an infinite cylinder with a single set of characteristic values of density, magnetic field strength and ionized fraction \footnote{We have verified that our results are almost insensitive to the chosen geometry by repeating the calculations for a spherically symmetric cloud.}.

The rate of contraction of a gas cloud is determined by the competing forces of gravity and magnetic pressure
\begin{eqnarray}
\label{em}
\frac{d^2r}{dt^2}=-2G\pi\rho r+ \frac{ B^2}{2\pi \rho r},
\end{eqnarray}
where $r$ and $\rho$ are, respectively, the characteristic radius and density of the cloud. The magnetic field evolves according to 
\begin{eqnarray}
\frac{dB}{dt}=\nabla \times (v_{\rm i}\times B)=-\frac{v_{\rm i}}{r} B,
\end{eqnarray}
where $v_{\rm i}$ is the velocity of the charged particles, which exceeds the velocity of the neutrals by
\begin{eqnarray}
\delta v=v_{\rm i}-\frac{dr}{dt}=\frac{(\nabla\times B)\times B}{4\pi x \gamma_{\rm D} \rho^2 },
\end{eqnarray}
where $\gamma_{\rm D}\approx 9\times 10^{13} \; {\rm cm^3s^{-1}g^{-1}}$ \cite{Os,DRD}.
The abundances of metals evolve according to 
\begin{eqnarray}
\label{ab}
\frac{d}{dt}\left(\frac{n_{\rm m}}{n_{\rm H}}\right)=-x_{\rm m}\frac{\delta v}{R}\left(\frac{n_{\rm m}}{n_{\rm H}}\right),
\end{eqnarray}
where $x_{\rm m}$ is the metal ionized fraction, $\Sigma x_{\rm m}n_{\rm m}=xn$. 

We assume that initially the relative abundances of metals are solar.
 Further we assume that $x$ scales with density as a power law, $x\propto \rho^\beta$, and that all ions are made up by metals with lowest ionization potential. Setting the time-scale for collapse of protostellar clouds, $t_{\rm coll}$, which is generally estimated to be of order 10 Myrs (e.g. Mouschovias et al. 2006), allows us to chose the initial value of $x$, which is almost directly proportional to $t_{\rm coll}$. Using these initial conditions, we integrate the equations (\ref{em})-(\ref{ab}) for different metallicities. 

\begin{figure} 
\centering
\mbox{\psfig{figure=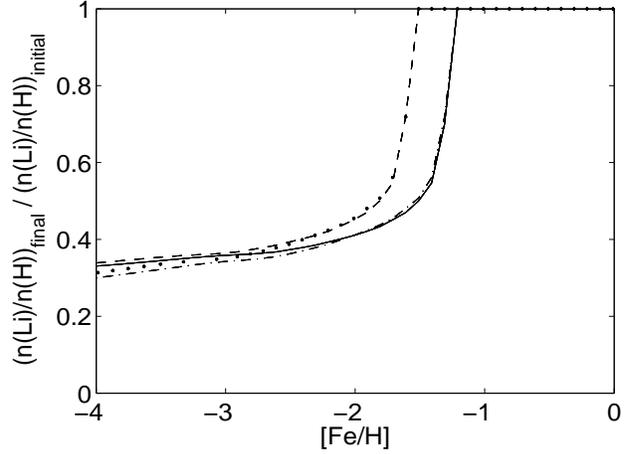,height=2.5in,width=3.5in}}
\caption{Lithium depletion as a function of stellar iron abundance. The solid and dashed lines are for $x\propto n^{-0.5}$ with $t_{\rm coll}=10$ and 5 Myrs, respectively.
The dot-dashed and dotted lines are for $x\propto n^{-0.6}$ with $t_{\rm coll}=10$ and 5 Myrs, respectively.}
\label{fig1}
\end{figure}

\begin{figure} 
\centering
\mbox{\psfig{figure=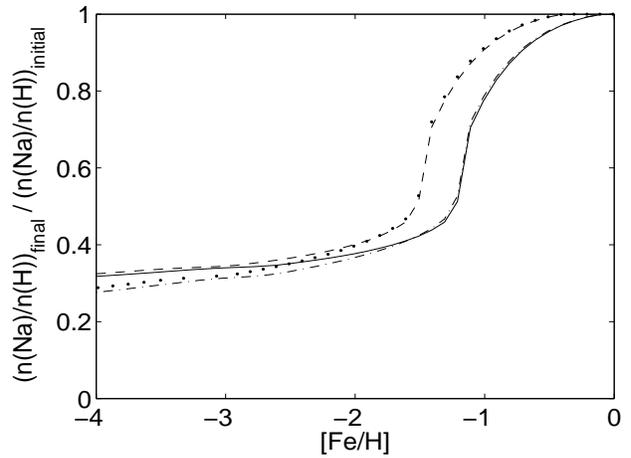,height=2.5in,width=3.5in}}
\caption{Same as Fig. \ref{fig1} for sodium. }
\label{fig2}
\end{figure}

\begin{figure} 
\centering
\mbox{\psfig{figure=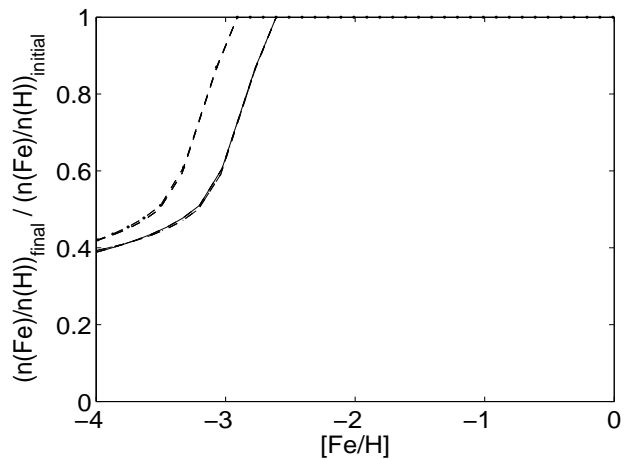,height=2.5in,width=3.5in}}
\caption{Same as Fig. \ref{fig1} for iron. }
\label{fig3}
\end{figure}

The results of the integration for lithium, sodium and iron abundances are illustrated by Figures 1-3.
At high metallicities the impact of ambipolar diffusion on element abundances is negligible, as the number of ions expelled by magnetic field is too small compared to the number of metal atoms. Increasing $t_{\rm coll}$, which allows to increase the number of expelled ions, or decreasing metallicity can make the impact of ambipolar diffusion more significant, but the depletion does not generally exceed a factor of a few. This limitation results from the fact that when magnetic flux drops by a factor of a few (with the corresponding drop in metal ion abundances) it can no longer effectively slow down the gravitational collapse and the dynamical time begins to fall much faster than ambipolar diffusion time-scale. Thus, at low metallicities the impact of ambipolar diffusion is only weakly sensitive to the assumed parameters ($\beta$ and  $t_{\rm coll}$), while at high metallicities it is generally negligible. On the other hand, the metallicity at which the transition between the low and high metallicity regimes occurs is very sensitive to $t_{\rm coll}$.

Several uncertainties in our model should be pointed out. So far we have assumed that all metals are in the atomic phase, neglecting the depletion on dust, which is highly uncertain for low metallicity/high redshift regions \cite{Ma}. 
Since most of the dust particles are charged, they can be also be expelled from gas by magnetic pressure, which opens an additional way of reducing metal abundances \cite{CM}. 
Another uncertainty is the magnitude of the magnetic field at high redshifts. Observations indicate existence of a strong magnetic field already at $z>2$ \cite{Ath}. However, at even higher redshifts stars might be forming in gas clouds with dynamically weak magnetic fields. Recently, Silk \& Langer (2006) have shown that even weak magnetic field would be sufficiently amplified during cloud collapse to become dynamically important, but in such a scenario the impact of ambipolar diffusion on metal abundances is very hard to estimate.

\section{Summary}
We have shown that in metal-poor clouds with dynamically important magnetic field the process of ambipolar diffusion results in lithium depletion by a factor of a few, which allows to reconcile WMAP measurements with the observed abundances. Likewise, the depletion of sodium may explain why [Na/Fe] ratio is constant in high metallicity stars ([Fe/H]$\gsim -1$), where the impact of ambipolar diffusion on the abundances is weak, but drops in the metal-poor ones \cite{CGS}. It should be noted, however, that since, unlike lithium, metals (such as Fe and Na) are not of primordial origin, their abundance patterns may have other causes besides diffusion.

In the metal-rich clouds the impact of ambipolar diffusion on the lithium abundance is negligible. However, by reducing the lithium abundance in the metal-poor stars, ambipolar diffusion is increasing its abundance in the inter-stellar medium. Therefore, later (i.e. metal-rich) generations of stars may in fact form with increased lithium abundance. Since the magnitude of this increase depends strongly on the previous star-formation history and the inter-stellar gas dynamics, there should be a significant scatter in the lithium abundance of the metal-rich stars, which is consistent with observations.

\section*{Acknowledgment}

This work was supported by the W.J. McDonald Fellowship of the McDonald Observatory.

\end{document}